\documentclass[paper]{JHEP3} % 10pt is ignored!

\usepackage{epsfig,multicol}

\def\be{\begin{equation}}
\def\ee{\end{equation}}
\def\beq{\begin{equation}}
\def\eeq{\end{equation}}

\def\bea{\begin{eqnarray}}
\def\eea{\end{eqnarray}}
\def\ds{\displaystyle}

\newcommand{\nn}{\nonumber}
\newcommand{\Lam}{\Lambda}
\newcommand{\eps}{\epsilon}

\title{ Renormalon disappearance 
in Borel sum of the $1/N$ expansion of the 
Gross-Neveu model mass gap} 

\author{ J.-L. Kneur and D. Reynaud\\
Physique Math\'ematique et Th\'eorique, UMR-5825-CNRS, \\
Universit\'e Montpellier II, F--34095 Montpellier Cedex 5, France}

\received{March 7, 2002} 		%%
\preprint{\hepth{0111120}}

\abstract{The exact mass gap of the $O(N)$ Gross-Neveu model is known, for
arbitrary $N$, from non-perturbative methods.
However, a ``naive" perturbative expansion of the pole mass exhibits
an infinite set of infrared renormalons at order $1/N$, formally similar to 
the QCD heavy quark pole mass renormalons, potentially leading to large
${\cal O}(\Lambda)$ perturbative ambiguities.
We examine the precise vanishing
mechanism of such infrared renormalons, which avoids this (only apparent)
contradiction, and operates without need of (Borel) summation contour 
prescription, usually preventing unambiguous separation
of perturbative contributions. As a consequence we
stress the direct Borel summability of the ($1/N$) 
perturbative expansion of 
the mass gap. We briefly speculate on a possible
similar behaviour of analogous non-perturbative  QCD quantities.}
\begin{document} 
\section{Introduction}
The $(1+1)$ dimensional $O(N)$ Gross-Neveu 
(GN) model\cite{gn,gnnext} often serves
as a simpler toy model for more complicated theories like QCD, sharing with 
it the 
properties of asymptotic freedom and dynamical mass generation, while
being an integrable model with many exact results available.
The exact mass gap (associated with the breaking of
the discrete chiral symmetry),
for arbitrary $N$, has been
evaluated from non-perturbative (NP) methods, more precisely from  
exact S matrix results\cite{exactS} associated with 
Thermodynamic Bethe Ansatz (TBA) methods\cite{TBAGN}. Yet
from a more general viewpoint, it can be  
an interesting issue (since still not fully clarified, in our opinion)
to study the precise matching between those 
two-dimensional NP exact results
on one side and the standard perturbative behaviour on the other
side. This may give some more insight on the short/long
distance physics interplay for more involved theories 
like four-dimensional QCD. 
In particular one apparent puzzle arises, once
realizing, as we examine here, that the 
(naive) perturbative expansion of the 
pole mass of the (massive)
GN model suffers at  next-to-leading $1/N$ order from  potential 
ambiguities, due to the presence of severe infrared renormalons,
which are indeed formally completely similar to the 
quark pole mass renormalons\cite{MPren}. While the TBA
GN mass gap\cite{TBAGN}, and a fortiori its next-to-leading $1/N$ 
expansion\cite{GN1N}, are unambiguously determined. 
Actually, independently of TBA results, it {\em is} expected on 
general grounds that any truly NP calculation should be free of such
ambiguities\cite{renormalons}, i.e. that infrared renormalons are perturbative
artifacts. But up to now only a few NP results have been explored
from this perspective, even for integrable models, thus 
we find instructive to
examine in some details how exactly the NP contributions,
here fully controllable at least at $1/N$ order, organize
themselves so that the necessary
vanishing of such ambiguities (indeed an infinite series of
ambiguities) occurs. 
Note that, what is in fact generally expected 
(and illustrated in a few explicit calculations 
in the (1+1) dimensional sigma
model\cite{David,BenBraKi}) is that the complete NP
result is unambiguous, but the separation of its perturbative from its NP
``operator product expansion" (OPE) contribution, is not. More precisely the
two contributions are intrinsically related through the need of a definite
prescription in choosing the integration path in the Borel plane,
if using e.g. Borel resummation methods, necessary
to avoid the renormalons in both parts and to resum the otherwise
ill-defined factorially divergent perturbative series. Now in
contrast, as we will show, a more interesting aspect of our results is that no
such prescription is needed for the NP infrared renormalon disappearance in
the GN mass gap.  More precisely, the
mechanism is such that the perturbative (re)expansion (to be
specified) of the pole mass at order $1/N$ defines a {\em directly} Borel
summable new perturbative series, which can be thus unambiguously separated
from the NP part. \\

The paper is organized as follows: in section 2, 
briefly recalling some known basic results of the
GN calculations at next-to-leading $1/N$ order,  
we shall define more precisely our ``naive" 
perturbative expansion of the $1/N$ order 
GN pole mass and exhibit
its infrared renormalon divergences.
In order to examine those issues, 
we will consider first the GN model with a non-zero
mass term, and single out a contribution which corresponds to a
purely perturbative information.
(The ordinary perturbative expansion of the true mass gap 
being of 
course zero to all orders, since the mass gap is non-perturbative
and of order $\Lam \sim e^{-2\pi/(N\,g)}$ where $g$ is the coupling). 
One can consistently recover the true mass gap in the massless
Lagrangian limit.  
In section 3, we rederive the known \cite{CampRoss,TBAGN}
exact calculation of the GN mass gap at $1/N$ order, 
but in the light of the
results of section 2. Namely, we stress the differences between
this exact $1/N$ calculation, which automatically takes into
account the full NP contributions, and the naive
pole mass expansion of section 2 based on a 
purely perturbative
information. We then perform in Section 4 a detailed calculation,
using a convenient (Mellin transform) method, to exhibit
the precise interplay mechanism between purely 
perturbative and NP contributions within the $1/N$ expansion,
and leading to the expected absence of renormalons in the
complete NP mass gap. We give in section 5 conclusion and some 
more speculative comments on a possible generalization
of those results to QCD. Finally, two short appendices are
devoted to additional technical details.    
\section{Perturbative renormalons in the $1/N$ naive pole mass}  
We start by briefly recalling  the
standard construction\cite{CampRoss,TBAGN} of the mass gap of 
the $O(2N)$~\footnote{From now on, all expressions correspond to the 
$O(2N)$ model, for easiest comparison with the exact\cite{TBAGN} and
$1/N$\cite{GN1N} mass gap results.} GN model  
at order
$1/N$, with a slightly different approach.  Here we consider in fact the model
with the addition of a Lagrangian mass term, in order to define the pole mass
in a somewhat more universal manner, making in particular the link
with analogous quantities in other models (QCD typically) more transparent. 
As usual the true
mass gap is to be considered in the chiral, massless Lagrangian limit.   The
Lagrangian thus reads 
\bea
{\cal L}_{GN} = \overline{\Psi} i \partial \hspace{-0.2cm} / 
\Psi - m \overline{\Psi}\Psi + \frac{g}{2} (\overline{\Psi}\Psi)^2
\label{GNlag}
\eea
and
the graphs
contributing to the two-point function at next-to-leading $1/N$ order are shown
in Fig. 1. 
At this $1/N$ order,
the renormalization procedure is relatively simple, since only 
the mass term is renormalized, which 
can be most simply performed by subtraction of the most
divergent terms in a taylor expansion of the integrands\cite{CampRoss}.  
%%%%%%%%%%%%%%%%%
%  FIGURE 1
\begin{figure}[htb]
\vspace{2cm}
\hspace{3cm}
\mbox{ 
\psfig{figure=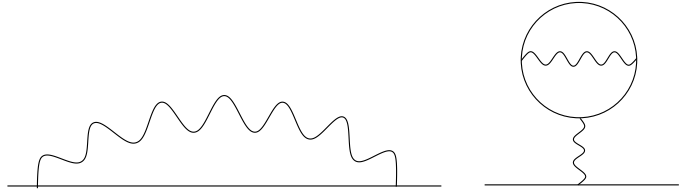,width=2truecm}}
\caption[]{The GN mass graphs at order $1/N$.}
\end{figure}
%%%%%%%%%%
The expression corresponding to Fig.1 reads, after angular integration:
\bea
M_P = M \left[ 1+ \frac{r_1}{N}+O\left(\frac{1}{N^2}\right)\right] 
\label{MPdef}
\eea
with $M = \Lambda \equiv 
\mu e^{-2\pi/(N g(\mu))}$ the mass
gap at leading $1/N$ order ($\mu$ is
the arbitrary renormalization scale), and   
\bea
\label{GNmass}
r_1 &= &\frac{g N}{4\pi\,M} \left[\int_0^\infty \! dq^2 (\frac{\;\not p}
{2p^2} [1- \frac{q^2+p^2+M^2_P}{A}] 
 - \frac{M_P}{A} ) G[q^2] \right. \\ \nonumber 
& & \left. -4 G[0] M_P\int_0^\infty \!\frac{dq^2}{q^2\,\zeta}
\ln[\frac{\zeta+1}{\zeta-1}]   \;G[q^2] \right] 
\eea
with $A \equiv \sqrt{(q^2+p^2+M^2_P)^2-4p^2\,M^2_P}$
($q$ is Euclidean but $p$ Minkowskian), 
$\zeta \equiv (1+4 M^2_P/q^2)^{1/2}$, 
and the dressed propagator
(wavy line) is    
\be
\label{Gq2}
G[q^2] = \left[1+ \frac{g N}{2\pi}[\ln\frac{M^2_P}{\mu^2} +\zeta
\ln[\frac{\zeta+1}{\zeta -1}]\:] \right]^{-1} 
= \left[\frac{g N}{2\pi}\zeta \ln[\frac{\zeta+1}{\zeta -1}] \right]^{-1} 
+{\cal O}(1/N)\;.
\ee
To obtain the pole mass at the next-to-leading $1/N$ order, and in the
massless limit, it is sufficient  
 to replace $\;\not \! p = M_P = M
[1+{\cal O}(1/N)]$ in Eq.(\ref{GNmass}),(\ref{Gq2})
up to $1/N^2$ terms. 
This reads
\bea
r_1 = \frac{1}{4} \int^{\mu^2}_0 \frac{d q^2}{M^2} (1-\zeta)\; \left[
\zeta \ln[\frac{\zeta+1}{\zeta -1}] \right]^{-1} \;
- \frac{1}{2} \int^{\mu^2}_0 \frac{d q^2}{q^2 \zeta^2} 
+{\cal O}(1/N)\;\;-subtraction
\label{r1pole}
\eea
where
a ``factorization" scale $\mu$
regularizing the integrals 
was introduced. The renormalization by subtraction
will remove the divergent part of integrands, 
also regularized in term of $\mu$. This
sharp momentum cut-off procedure is rather similar to the one in 
ref.~\cite{GN1N}, giving results
perfectly consistent with these authors, as we will see in next
section, and is convenient for our purpose here, where we
are primarily interested in the asymptotic
behaviour of (\ref{r1pole}) (thus for $\mu \gg \Lam$, but kept finite). \\

For simplicity we let aside in this section the tadpole second 
graph in Fig. 1, 
as its exact expression in (\ref{r1pole}) does not give any
factorially divergent perturbative coefficients when expanded\footnote{Due
to the second equality in (\ref{Gq2}), valid at this $1/N$
order, which makes the tadpole graph simplifying
to purely polynomial integrals. Beyond the $1/N$ order, one could still
in principle choose the arbitrary scale $\mu$ such that (\ref{Gq2}) holds.}. 
[It gives however a finite contribution to the mass gap, as we shall see in
next section when a precise evaluation of the mass gap and its
asymptotic behaviour will be considered.]  
We shall next examine in some detail how a naive perturbative expansion 
of (\ref{r1pole})
leads to infrared renormalons and 
related ${\cal O}(\Lambda/\mu)$ ambiguities. 
Before to proceed to see that,
 it is perhaps
important to better explain
what we exactly mean by ``naive" perturbative expansion
of  (\ref{r1pole}). Clearly, the expression (\ref{r1pole})
is exact and complete at $1/N$ order, which means implicitly
that it contains a priori {\em both} perturbative and 
NP contributions. One may argue that, in fact, 
the ordinary perturbative expansion of the true mass gap is 
zero to all orders, the mass gap being a NP
quantity  of order $\Lam$. However, this is true strictly
speaking only when considering the massless
GN model, i.e. for $m\to 0$ in Eq. (\ref{GNlag}). Alternatively,
one may consider the calculation at this $1/N$ order
of Eq. (\ref{r1pole}) as valid for the massive GN Lagrangian case where 
say, $m \sim M_P ={\cal O}(\Lam)$ (for completeness, the case where 
$m \gg \Lam$ is also briefly discussed in Appendix A). 
Remark indeed that in (\ref{r1pole}), the $1-\zeta$
factor, which corresponds to the skeletal ``one-loop" 
integrand of the first graph
in Fig. 1, is essentially obtained from the pole mass
condition $\;\not \!\! p = M_P$, irrespectively of whether $M_P$
is ${\cal O}(\Lam)$ or not. 
The remaining $\zeta$--dependence is the dressed scalar propagator, 
as defined in Eq. (\ref{Gq2}), where the leading $1/N$ order mass
gap definition is only used in the right-hand side of 
Eq. (\ref{Gq2}).
Thus the perturbative form of (\ref{r1pert}) is quite
generic, formally
also applying to e.g. the QCD pole quark mass [but with obviously a
different $q^2$--dependence replacing the $1-\zeta$ term]. 
Now, since in the present case
one can control everything at the 
$1/N$ order explicitly,
our purpose is to consider the kind of results obtained when   
the NP knowledge is partially omitted. More precisely, let us assume
that within the next-to-leading $1/N$ graph structure
in Fig. 1, only the 
purely perturbative part of the     
dressed scalar propagator is known. Our motivation
is that this is analogous to the
situation in a more complicated theory like QCD typically. 
Starting thus from the 
``purely perturbative"
information means replacing in 
(\ref{r1pole}) the dressed scalar propagator 
by the effective coupling\footnote{In the sequel
we rescale $b_0\, g \equiv N/(2\pi)\:g \to g$, to define the $1/N$
expansion properly, and absorbing as well the $2\pi$ factor just for
convenience.}: 
\be
r_1 = \frac{1}{4} \int^{\mu^2}_0 \frac{d q^2}{M^2} (1-\zeta)\; 
\left[g(q^2) -2g\,\frac{M^2}{q^2} +{\cal O}(\frac{M^2}{q^2})^2 \right]
\label{r1pert}
\ee
where in the bracket of Eq. (\ref{r1pert}) the ${\cal O}(M^2/q^2)$ ``NP"
corrections to the effective coupling are of course exactly known in the GN
case and can be obtained explicitly 
here from a systematic expansion of the right-hand side of  
Eq.~(\ref{Gq2}) for large $q^2$. Again, we stress that omitting this
NP part is evidently not fully consistent, but the purpose of this
(somewhat artificial) separation of the different contributions 
is to examine precisely how the renormalons
(would) appear, in a way very similar to the QCD pole mass, 
and more importantly from which detailed
mechanism do they finally disappear in the GN case,  
once the complete NP information is taken
into account. \\
The simplest standard
procedure to exhibit the IR renormalons is by expanding the
$1-\zeta$ term for small $q^2$: 
\be
1-\zeta = 1-\frac{2M}{q}\;(1+\frac{q^2}{4M^2})^{1/2}
\simeq 1-\frac{2M}{q} +{\cal O}(\frac{q}{2M})
\label{zeta}
\ee
and expanding the effective coupling $g(q^2)$ 
at one loop order of renormalization group (RG) in powers of
$b_0 g(\mu) \ln [q^2/\mu^2]$. From (\ref{zeta}) it is seen that the leading
singularity comes from the $q^{-1}$ term which, combined with the  $(g(\mu)
\ln [q^2/\mu^2])^n$ terms of the effective coupling expansion, produces
factorially divergent perturbative coefficients:
\be
r_1^{leading} \sim -\frac{\mu}{M}\;\sum^\infty_{p=0} 2^p p! \; g^{p+1}(\mu)\;. 
\label{leadren}
\ee
The non sign-alternation of those factorial coefficients
implies\cite{renormalons}
that the corresponding series is not Borel summable:
the Borel integral corresponding to (\ref{leadren}) reads 
\be
BI[r_1^{leading}] \sim  -\frac{\mu}{M}\;\int^\infty_0 dt e^{-t/g} (1-2t)^{-1}  
\label{BIlead}
\ee
so that the pole at $t_0 =1/2$  on the integration path gives the leading
ambiguity for the pole mass 
\be
\delta M_{leading} \sim
\lim_{\epsilon \to 0} [\int^{\infty+i \epsilon}
-\int^{\infty-i \epsilon}]
(dt\: e^{-t/g} (1-2t)^{-1}) = \pm  i\pi \mu e^{-t_0/g}  \propto \Lambda \;.
\label{leadambig}
\ee     
Note that this leading renormalon ambiguity of ${\cal O}(\Lambda)$ is
completely similar to the one derived for the quark pole mass in
QCD\cite{MPren}. 
The fact that the perturbative contribution Eq. (\ref{BIlead})
is proportional to the factorization scale $\mu$ may appear
at first puzzling (since it is supposed to be a contribution
to the mass gap $\sim \Lam \ll \mu$), it is in fact also
a known feature of the non Borel summability of
the corresponding series\cite{renormalons} with this explicit
cutoff regularization:
namely, it indicates us that an additional  NP contribution, also
proportionnal to the factorization scale, is necessarily needed
(at least in this regularization scheme) 
in order to recover the right mass gap dependence $\sim \Lam$.\\
Although our above derivation (starting from
purely perturbative information) is very standard, 
the fact that the GN model perturbative {\em pole} mass at
$1/N$ order also has the specific
structure (\ref{leadren})--(\ref{leadambig}) of {\em infrared} renormalons,
was perhaps not clearly appreciated before, to our
knowledge\footnote{In ref.\cite{gn} appeared already
the (earliest) discussion
on {\em ultraviolet} renormalons
in a field theory framework, which were shown to be harmless
(Borel summable) for the asymptotically free GN model.}. 
Similarly, we can easily check from (\ref{GNmass}) 
that the leading asymptotic behaviour of
perturbative coefficients of the two-point function
for $p^2 \gg M^2$ is $\sim \sum_p  p! \; g^p$, which
accordingly gives a less severe ambiguity  of ${\cal O}(\Lambda^2/p^2)$,
where again one can note the similarity with
the QCD off-shell $p^2 \gg M^2$ quark correlation function
case\cite{MPren,renormalons}.\\ It is in fact possible to
go a step further and to characterize at arbitrary next orders the
renormalon
properties of the naive perturbative expansion of the $1/N$ GN mass gap.
Consider the second order (exact) RG dependence of the
effective coupling\cite{gn2}:
\be
g/g(q^2) \equiv f = 1+b_0 g \ln \frac{q^2}{\mu^2}
+\frac{b_1}{b_0}g \ln \left[ f \frac{(1+b_1/b_0 g f^{-1})}{1+b_1/b_0 g}
\right] 
\label{fdef}
\ee
defining $f$ recursively, with $g \equiv g(\mu)$
and the beta function   
$\beta(g) = -2b_0 g^2
-2b_1 g^3 -\cdots$ (where for clarity we reintroduce the original
coupling and RG coefficients, i.e. before rescaling of $g$). 
We can put  
Eq.~(\ref{r1pole}), with (\ref{fdef}), directly into the form of a Borel
integral,  after a convenient  change of variable\cite{Grunberg}, defining the
(Borel) variable $t$ as 
\be
b_0\:t = \frac{1-f}{1+b_1/b_0 g}\;.
\label{deftf}
\ee
Taking expression (\ref{r1pert}), but using now Eq.(\ref{fdef}), 
we find after
some algebra 
\bea	
\label{RG2}
&r^{RG2}_1  = &\ds -\frac{\mu}{2M} \int^\infty_0 dt
e^{-\frac{\alpha t}{2}}  (1-b_0 t)^{-1- C} \; 
\left[1 -\frac{\mu}{2M} e^{-\frac{\alpha t}{2}} (1-b_0
t)^{-C} \right.\\ \nonumber 
&  &\left. + 
\frac{\sqrt \pi}{2}
\sum^\infty_{p=1}  \frac{(\mu^2/4M^2)^p}{p! \Gamma[3/2-p]}\; e^{-p \alpha t}
(1-b_0 t)^{-2 p C} 
\right]
\eea 
where $C = b_1/(2b^2_0)$ and $\alpha= 1/g+b_1/b_0$. To obtain
(\ref{RG2}) we expanded Eq.(\ref{zeta}) in powers of  $q^2/M^2$,
and used Eqs. (\ref{fdef},\ref{deftf}). This gives the complete 
(leading and all subleading orders) series of
infrared renormalons (initially corresponding to cuts at 
$b_0 t_p=
1/2,1,..(2p+1)/2$, $p \in \mathbf{N}^*$). The change of variable
(\ref{deftf})  makes calculations more convenient since expression (\ref{RG2}) 
has only a cut at $t \ge 1/b_0$. Now 
we can calculate the ambiguity to all 
(perturbative) orders (of course still limited
to $1/N$ order), which we define by the difference
of contour above and below the cut. We find, again after
some algebra:  
\bea
\label{fullambig}
&(2\pi i)^{-1} \delta M^P = & \pm \ds \frac{\bar\Lambda}{2 b_0}
\left[\frac{(2e)^{-C}}{\Gamma[1+C]} -\frac{\bar\Lambda}{2M}
\frac{e^{-2C}}{\Gamma[1+2C]} \right. \\ \nonumber  
&  & \left. + \frac{\sqrt
\pi}{2} (2e)^{-C} \sum^\infty_{p=1} \frac{(2p+1)^{(2p+1)C}}
{p! \Gamma[3/2-p] \Gamma[1+(2p+1)C]}
(\frac{(2e)^{-C}\:\bar\Lambda}{2M})^{2p} 
\right]
\eea
where we used essentially
\be
\lim_{\epsilon \to 0} [\int^{\infty+i \epsilon}_0 - \int^{\infty-i
\epsilon}_0] dt e^{-\alpha t} (1-\beta t)^\gamma = 2\pi i \;e^{-\alpha/\beta}
\beta^\gamma \alpha^{-(1+\gamma)} \Gamma[-\gamma]^{-1}
\label{contint}
\ee 
and identified the $\overline{MS}$ scale 
$\bar \Lambda = \mu e^{-1/(2b_0g)}
(b_0 g)^{-C} [1+(b_1/b_0) g]^C $
consistently at second RG order.
Eq.(\ref{fullambig}) thus gives the full series of ambiguities due to infrared
renormalons (for the first graph in Fig.~1), in the form of power corrections
of order $\Lambda^p$, with the first term in the bracket
of (\ref{fullambig}) the leading order ambiguity of ${\cal O}(\Lam)$. 
If $b_1
=0$ (e.g. at first RG order) expression inside the bracket of Eq.
(\ref{leadambig})   simplifies to $(1+r^2/4)^{1/2}-r/2$, $r\equiv \Lam/M$
(indeed sufficient at this $1/N$ order, since from 
Eq.~(\ref{MPdef}) $r_1$ is already the $1/N$ term.)\\
As already emphasized, the above derivation of (\ref{fullambig}) only uses
information that is in fact purely perturbative: the effective coupling
at second RG order, Eq. (\ref{fdef}), and the specific GN mass $q^2$
kinematic dependence,  Eq.~(\ref{zeta}) of
the  (one-loop) skeletal first graph in Fig.~1. Accordingly, 
a similar derivation is 
possible for the QCD quark pole mass
renormalon properties (indeed the 
equivalent of the 
information 
in (\ref{fullambig}) in the QCD case is also known, though perhaps
expressed in a slightly different form\cite{renormalons}).\\
Actually, by considering only the first graph of Fig.~1 we slightly
oversimplified the complete renormalon picture for the GN model: clearly, the
second tadpole graph also gives renormalons, {\em if} considered purely
perturbatively. These are easily analyzed similarly to 
Eqs.~(\ref{r1pert})-(\ref{zeta}), and
the expanded integrand for $q^2\to 0$ leads to renormalon poles at $b_0
t_p  = p \in \mathbf{N}^*$. Incidentally, the pole at $b_0 t=1$ exactly
cancels the one in (\ref{RG2}), but the leading as well as all subleading poles
for $b_0 t_p \ge 3/2$ in (\ref{RG2}) remain uncancelled.\\
In summary, we thus observe from the structure of (\ref{fullambig}) that even 
resumming the full integrand $1-\zeta$
(taking the full series of sub-leading
renormalons) does not remove in any way the leading ambiguities (even if there
are some "accidental" cancellations among subleading poles at this purely
perturbative level, between the two graphs of Fig. 1 as above discussed). 
This is not at all surprising, 
since as emphasized  (\ref{RG2}),(\ref{fullambig})
are still  perturbative calculations.
This ambiguity in the purely perturbative
piece of the pole mass is severe, being of the same ${\cal O}(\Lam)$
order than the mass itself, similarly to the QCD pole
mass\cite{MPren} ambiguity. 
As explained above in detail, this originates from
the $1-\zeta$ factor, specific to the pole mass, which has an expansion for
$q^2\to 0$ whose leading term is ${\cal O}(|q|^{-1})$, cf. Eq. (\ref{zeta}). 
It is interesting to compare those pole mass results with what happens
for other related quantities of interest.  For instance, the GN
fermion  condensate $\langle \bar \Psi \Psi \rangle$ (which is related
to the mass gap, since e.g.  $\langle \bar \Psi \Psi \rangle = N
M_P$ in the $N \to \infty$ limit), has a leading ambiguity 
in its perturbative expansion which is  
much power suppressed. More precisely, the relevant part of the
condensate can be obtained, roughly speaking, by closing the external
leg of the first graph in Fig. 1. A straightforward calculation
along the lines of Eqs. (\ref{r1pert})--(\ref{leadambig}) gives for
the relevant integral:
\be
\langle \bar \Psi \Psi \rangle \sim 
\int^{\mu^2}_0 dq^2 [c_0 +c_1 \ln M +{\cal O}(\frac{q^2}{M^2})\; ]
\;G[q^2]
\label{ffbar}
\ee
where $c_0$ and $c_1$ are unessential constants not specified here.
Accordingly, when takin in   
(\ref{ffbar}) only the perturbative part $g(q^2)$ of the propagator
$G[q^2]$, cf. Eq.~(\ref{r1pert}),
one finds for the Borel transform an IR renormalon
pole at $b_0 t =1$, thus associated with an ambiguity of
${\cal O}(\Lam/\mu)^2$. 
This result for the IR renormalon in the perturbative tail of the GN condensate
was in fact obtained long ago\cite{DavHam}, both in this
heuristic form and rigorously by
using a lattice regularization with Wilson fermions, where it was
precisely shown that this absence of a leading order 
renormalon is due to the
chiral symmetry. As concluded by these authors, even if the
perturbative expansion of the condensate is not, strictly speaking, 
Borel summable, the ambiguity
may be neglected since it is power suppressed with respect to the NP condensate
itself, the latter being of  ${\cal O}(\Lam)$. 
A similar suppression due to chiral symmetry of the leading IR renormalon
also occurs for the QCD quark condensate in four dimensions.\\
Coming back to the GN 
pole mass case, we have however to stress once more at this
point that all the  perturbative renormalon results derived in this
section are  in a sense artifacts, since the perturbative expansion that we
considered was on purpose artificially generated as an 
incomplete part of the full $1/N$ expression of the mass gap.
\section{Borel summability of the exact ${\cal O}(1/N)$ mass}
Alternatively, since expression (\ref{r1pole}) is exact at $1/N$ order,
we can
calculate exactly the expression of the mass gap,
i.e. without truncating (\ref{r1pole}) to its perturbative expansion. 
Indeed integral (\ref{r1pole}) can
be evaluated analytically exactly: after a convenient change of
variable $\zeta^{-1} = \tanh(\frac{\phi}{2})$
we obtain
\be
r_1 = \frac{1}{2} \left[ Ei[-\theta] -\ln \theta 
-\gamma_E +\ln(\ln \frac{\mu^2}{M^2}) -2 \ln (\cosh[\theta/2])+\ln
\frac{\mu^2}{M^2}\;\right] \label{MexactEi}
\ee
with $\chi=(1+4 M^2/\mu^2)^{1/2} \equiv 1/\tanh(\theta/2)$ (i.e.
$\theta = \ln[(\chi+1)/(\chi-1)]\;\ge 0$), 
and $Ei(-x) \equiv -\int^\infty_x dt e^{-t}/t$ ($x \ge 0$) the Exponential
Integral function. The term $-2 \ln (\cosh[\theta/2])$ in (\ref{MexactEi})
corresponds to  the (unsubtracted) tadpole graph of Fig.1, and the terms 
$\ln\ln (\mu^2/M^2) \equiv -\ln g $ and $\ln \mu^2/M^2 \equiv 1/g$  
are the subtraction terms for the first and tadpole graphs, respectively.
One can easily check the finiteness of (\ref{MexactEi}), if letting
the ``cutoff" $\mu \to \infty$. Now, 
as already stressed, here we are interested in the complete
asymptotic behaviour, thus letting $\mu \gg \Lam$ but kept finite, retaining
eventually all the power correction terms in $\Lam/\mu$.
The function
$Ei(-x)$ for $x >0$ has an asymptotic expansion with factorial but {\em
sign alternating} coefficients, therefore explicitly Borel summable 
and perturbatively unambiguous. More precisely
when re-expanding (\ref{MexactEi}) in perturbation, we obtain
\be
M^P = M \left[1+ \frac{1}{2N} \left( 2\ln 2 -\gamma_E  
-\frac{M^2}{\mu^2}  [\sum^\infty_{n=0} (-1)^{n} n!\;g^{n+1}\; +{\cal O}
(\frac{M^2}{\mu^2})]
\right) \right] 
\label{genuine}
\ee
where the higher order power correction terms, that we do not give 
explicitly here, can be
obtained  by a systematic expansion in $M^2/\mu^2$ of $\theta$.
The tadpole graph 
in Fig.1 contributes
a constant term $2\ln 2$, relevant to
the mass gap determination, but 
does not contribute the factorial asymptotic behaviour of the perturbative
series, as already mentioned. \\
We emphasize that the specific $1-\zeta$ form of the integrand in
(\ref{r1pole}) plays an essential role for the Borel summability of expression
(\ref{genuine}), which is accordingly peculiar to the pole mass. In
contrast, the off-shell two-point function
expression (\ref{GNmass}) for arbitrary $p^2$ may be evaluated similarly
non-perturbatively (at $1/N$ order),  
and does not lead to a directly
Borel summable perturbative series, in consistency with the results
obtained and discussed previously in 
ref.~\cite{CampRoss}\footnote{\label {NBS}
Technically, the simplest such terms can be expressed in terms of 
$Ei(\theta)$,
which accordingly (since $\theta >0$)
has asymptotic expansion with same sign factorial coefficients
and imaginary part $\pm i \pi$.}. 
We shall come back to this specificity of the pole mass 
in more details in next section 4. 

We also stress that the use of the ${\cal
O}(1/N)$ mass gap relation Eq.(\ref{MPdef}) in the Lagrangian 
mass $m\to 0$ limit,
though it greatly simplifies the analytic 
evaluation of Eq.(\ref{r1pole}), plays no
particular role in the good 
asymptotic properties of Eq.(\ref{genuine}): more
precisely, starting from
the exact $1/N$ mass expression (\ref{r1pole}), and introducing the 
arbitrary mass dependence $m$, Borel summability is maintained
with an asymptotic expansion similar to Eq.(\ref{genuine}) for any value of
the pole mass  $M_P \gg \Lam $ (see Appendix A).\\    
Finally to obtain the correct mass gap
$M^P/\bar \Lam$ at next-to-leading $1/N$ order from expression
(\ref{MexactEi}), one should yet introduce the $\overline{MS}$ basic scale
above defined after Eq. (\ref{contint}). Dropping terms of higher ${\cal
O}(1/N^2)$ order, we obtain \be
M^P/\overline{\Lam} = 1+\frac{r_1}{N}+{\cal O}(\frac{1}{N^2}) = 
1+\frac{1+2\ln 2 -\gamma_E}{2N}
\label{mgapN}
\ee
in agreement with ref.\cite{GN1N}. Note however that our expression
(\ref{MexactEi}) differs in fact from ref.\cite{GN1N}, more precisely
by the $Ei[-\theta]$ term.
This is simply because in ref.\cite{GN1N} 
all terms vanishing as inverse powers of $\mu$ were
dropped, which is sufficient to identify the mass gap Eq.~(\ref{mgapN}). 
From our result, this is consistent because at the
(non-perturbative) level of Eq.(\ref{MexactEi}), all those ``power correction"
contributions from $Ei(-\theta)$ can be unambiguously separated, thus
dropped from Eq.(\ref{genuine}), to let only the part relevant to determine
the mass gap. The explicit Borel summability
of the ``educated" perturbative expansion of the pole mass at $1/N$
order, Eq.~(\ref{genuine}), confirms
the consistency of the whole procedure. An equivalent signature 
of Borel summability here is the
absence of positive powers of $\mu$ in (\ref{genuine}), in
contrast with the purely perturbative contribution 
Eq.~(\ref{leadren}). 
Applying the same procedure to other
quantities than the pole mass, one eventually ends up with
asymptotic expansions with non sign-alternated factorial
coefficients and $\mu^P$ power terms with $P >0$
(as is clear from the above expression of $\theta$, cf.
footnote \ref{NBS}). 
Within a complete NP calculation,
the latter $\mu^P$ terms in fact
cancel when combining the OPE and perturbative parts, as 
illustrated in explicit calculations for some vacuum 
expectation value\cite{David,Novikov}
and off-shell correlation functions\cite{BenBraKi} 
of the $O(N)$ sigma model at
$1/N$ order.
As mentioned in introduction,
this also illustrates that, though
any complete NP calculation is expected to be renormalon 
ambiguity free, one cannot expect in general (for
an arbitrary Green function) to be able
to single out unambiguously the perturbative 
contributions.\\  
Coming back to our result Eq.~(\ref{genuine}), it appears 
thus immediately in
apparent contradiction with those obtained starting from purely pertubative
expansions, (\ref{leadren}--\ref{fullambig}) above. We shall examine in next
section how to reconcile these two different pictures.   
\section{Explicit disappearance of IR renormalons}
How exactly the ``bad" factorial coefficients with
no sign alternation in Eq. (\ref{leadren}) disappear, or more precisely 
transmute
into ``good" sign-alternated factorials, in the
exact expression (\ref{genuine})? Clearly, it can only 
be that this necessary
change of the wrong sign perturbative expansion factorial
coefficients should
occur with the NP power expansion contributions:
the weak point of the standard perturbative renormalon picture is that
we have expanded the $1-\zeta$ term in the infrared low $q^2$ regime, while
keeping the short distance, perturbative effective coupling form of the
propagator.
This is motivated, as already mentioned,  
from the fact that the latter information
is a priori the only accessible one in more involved theories such as QCD.
In the present case, as we know exactly the mass at $1/N$
order one may at first hope to infer such transformations by examining e.g. the
systematic short distance $q^2\to\infty$  and long distance $q^2\to 0$ power
expansion of the integrand in (\ref{r1pert}), which are perfectly 
well-defined\footnote{E.g. the infrared $q^2\to 0$
expansion of the dressed propagator is purely polynomial, with no log.
dependence: $[ \zeta \ln[\frac{\zeta+1}{\zeta -1}] ]^{-1}  
\simeq \frac{1}{2} -\frac{q^2}{24M^2} +\cdots$.}.\\
In fact, to see the vanishing of renormalons operating needs a little more sophisticated
analysis. Following e.g. refs.\cite{David,BenBraKi}, we introduce the 
Mellin-Barnes (MB)
transform for a part of the integral Eq.~(\ref{r1pole}), which then takes the
form (again omitting here the tadpole graph for simpler illustration): 
\be
r_1 = \frac{1}{4} \int^{\mu^2}_0 \frac{dq^2}{M^2} 
\int^\infty_0 dt \frac{1}{2\pi\: i} \oint ds K(s,t)
(\frac{M^2}{q^2})^{-s} 
\label{Mellin}
\ee  
where 
the Kernel (inverse MB transform) is defined in our case by
\be
K(s,t) =\int^\infty_0 dx x^{s-1} (1-\zeta)\;\zeta^{-1} 
 \left[\frac{\zeta-1}{\zeta +1}\right]^t\;.
\label{KMellin}
\ee
The
MB transform
method main purpose is that it will exhibit precisely the
singularities of the integrand, in the Borel plane $t$ of interest.
The sequel is just algebraic  manipulation. Changing again variable 
$\zeta^{-1} = \tanh(\frac{\phi}{2})$, {\em except} for the $1-\zeta$ term 
kept on purpose, using 
Eq. (\ref{zeta}), as an
expansion in $q/M$,
(\ref{KMellin}) can be evaluated exactly, and   
\be
\int^{\mu^2}_0  \frac{dq^2}{4 M^2} (\frac{M^2}{q^2})^{-s}\:K(s,t) =
\frac{\mu^2}{2M^2} \sum_{a \ge -1} 2^{-a} c_a\;
\frac{\Gamma[1+a-2s]\Gamma[-a/2+s+t]}{\Gamma[1+a/2 -s+t]\:(1+s)} 
(\frac{M^2}{\mu^2})^{-s}
\label{KMB}
\ee  
defined for $Re[s+t] >a/2$ and $Re[2s] <1+a$. The latter conditions are such
that integral (\ref{KMellin}) converges, and play essential role in
determining the singularities. The variable $a$ in (\ref{KMB}) is 
simply the power
of $q/M \sim \sinh \phi$ in  expansion (\ref{zeta}), with coefficient $c_a$
respectively. Thus $a=-1$ with $c_{-1}= -2$ corresponds to the leading
renormalon, and $a=0,..2p+1$, $p \in \mathbf{N}$ to subleading ones.
To evaluate the $s$ integral one
can simply close the contour on the left, 
 and sum over residues of the poles included
in this domain (since $x^{-s} \equiv (M^2/q^2)^{-s}$ decays
exponentially fast for the asymptotic regime $q^2 \gg M^2$ we are interested
in). Expression (\ref{KMB}) has (simple) poles at $s=-1$, $s=a/2-t-k$, 
and $2 s = a+1+ k$, $k \in \mathbf{N}$, where the latter poles do not
contribute for the relevant contour. The final result is 
\bea
\ds
r_1 &\sim &\frac{\mu^2}{4M^2}\int_0^\infty dt \sum_{a \ge -1} \frac{c_a}{2^a}
\left[\frac{\Gamma[3+a]\Gamma[t-1-a/2]}{\Gamma[2+t+a/2]} \frac{M^2} {\mu^2} 
\right. \nn \\ 
& & \left. +
e^{-t/g} \sum^\infty_{k=0} \frac{(-1)^k \Gamma[1+2t+2k]} {k!  \Gamma[1+2t+k]
(1+a/2-t-k)}(\frac{M^2}{\mu^2})^{k-a/2}  \; \right]  \label{MPresidue}
\eea
where the first term in bracket corresponds to the residue of the pole at
$s=-1$, while other terms correspond to the sum over residues of the poles at
$s=a/2-t-k$, and we used $ (M^2/\mu^2)^t \equiv e^{-t/g}$.
In (\ref{MPresidue}) one sees that the first term in the bracket and the
summed terms both have poles at $t(a,k) =1+a/2-k$, which can occur at $t>0$
depending on $a,k$ values. On more physical grounds, the contributions from the
first term, the initially $s=-1$ pole, originate from  
power terms $(M^2/q^2)^{-s}$ in Eq.(\ref{Mellin}), 
and correspond intuitively to 
non-perturbative "OPE" contributions, while the perturbative contributions 
are those multiplied by
$e^{-t/g}$ in (\ref{MPresidue}). 
Even though in fact both contributions are 
in the end of the same (non-perturbative) order, such as to compensate
the bad poles in $t>0$,
it is natural within the Borel transform formalism
to distinguish those two contributions
as ``non-perturbative" and ``perturbative" respectively,
since typically the second contribution has the characteristic $t$-dependent  
overall factor $e^{-t/g} = (M^2/\mu^2)^t$ multiplying
an expression with poles at $t>0$, so that when this latter
contribution is considered alone it organizes as an ordinary
perturbative expansion when integrated over $t$. More precisely, for example
the poles at $t =1/2$ for $a=-1$   correspond to the leading order renormalon,
with $k=0$. Indeed, keeping only the leading renormalon perturbative terms, 
$\propto e^{-t/g}$, for $k=0$, one recovers exactly Eq.~(\ref{BIlead}). \\  
Now it can be easily checked  that this
$t =1/2$ pole in fact cancels exactly against the first NP term 
$t =1/2$ pole, and similarly for all subleading poles at $t
=1,3/2,..(2p+1)/2$. This  is the announced disappearance of
renormalons. Moreover 
all cancellations
happen, for a given pole at $t =1+a/2-k$, 
between NP and perturbative terms of the same
$k$ values. Since (\ref{MPresidue}) is in the form of a Borel integral, and
after cancellations all remaining poles occur at $t<0$,
it defines a Borel summable series, whose leading terms just
correspond to the  asymptotic series  defined in (\ref{genuine}). 
To see that, it is simpler to alternatively proceed directly 
with (\ref{KMellin}) using the change of variable $\zeta \to \phi$ for any
terms in the integrand, i.e. without going through power expansion
(\ref{zeta}):  after MB transform one ends
up directly with a Borel integral where no poles at all appear at $t >0$,
and which exactly gives the asymptotic series in
Eq.(\ref{genuine}), including correct finite terms $-\gamma_E +.. $. 
[NB
there are also poles at $t=0$ in (\ref{MPresidue}), which as
usual\cite{renormalons} simply corresponds to the UV divergences, and are
removed consistently by the appropriate subtraction terms, that we do not
display explicitly.] 
For instance, illustrating only the terms from the first graph in Fig.1,
after cancellations of the $t >0$ poles the MB transformation gives
\bea
\ds
r_1 &\sim &-\frac{\mu^2}{2M^2}\int_0^\infty dt 
\left[\frac{M^2}{\mu^2}\frac{1}{t(1+t)} 
+e^{-t/g}[ -\frac{1}{t} \frac{M^2}{\mu^2} +\frac{3+2t}{1+t} 
(\frac{M^2}{\mu^2})^2 +{\cal O}(\frac{M^2}{\mu^2})^3\:]\right] \nn \\
&\sim &
-\frac{1}{2} [\gamma_E +\frac{M^2}{\mu^2} g \sum^\infty_{n=0}
(-1)^n n! g^n \:]
\label{MBresult}
\eea
to compare with Eq.(\ref{genuine}) (where we used e.g. 
$\int^\infty_0 dt\, t^{-1} [1/(t+1)-e^{-t}] \equiv \gamma_E$).  
Note that in Eq. (\ref{MBresult}) all the ``bad" positive powers
of $\mu/M$, that signalled the non-Borel summability in the corresponding
incomplete purely perturbative expressions Eqs.~(\ref{BIlead}), 
(\ref{RG2}), have now disappear. Of
course,  proceeding in this ``direct" way is nothing but a consistency check
that the MB transform gives a correct alternative calculation of the
asymptotic expansion of the exact integral Eq.(\ref{r1pole}), which we started
from anyway. But our explicit separation of the expanded perturbative
renormalon part (\ref{zeta}) in connection with the MB method allows to
visualize explicitly the renormalon disappearance order by order in
(\ref{MPresidue}). For completeness note that 
a very similar MB transform analysis can be performed for the tadpole graph
renormalons, which disappear similarly from a detailed analysis
that we do not display here.  

It may be of interest to examine the influence of the regularization
scheme on the results derived above. As expected,  the
renormalon properties are independent of the regularization
used, but they manifest themselves
in a quite different form e.g. in dimensional
regularization\cite{David,renormalons}.
The momentum cutoff regularization prescription 
considered here, introducing
an explicit factorization scale $\mu$, cf. Eq.
(\ref{r1pole}), is clearly very convenient for our purpose since
it directly displays the NP power correction terms, while
the interpretation of the latter 
for instance in dimensional regularization (DR) is not completelly
obvious. Nevertheless, the appearance (and disappearance)
of IR renormalons as discussed here can also be seen equivalently in
DR. 
As is known from general analysis\cite{David,renormalons}, 
the IR renormalon
ambiguities in DR manisfest themselves in the fact that the
Borel integral over the perturbative
contributions has to be evaluated 
for complex $\eps$, and gives a different result depending
on whether $Im[\eps]>0$ or  $Im[\eps]<0$. The difference between
the two results is proportional to the NP contribution. 
In a rather simplified analysis of the GN pole mass (see Appendix
B for more detail), 
considering the equivalent of the MB plus Borel transform
Eqs. (\ref{Mellin})--(\ref{MPresidue}) in DR, one arrives
at an expression quite similar to Eq. (\ref{MPresidue}),   
but in which now both terms
have poles at $t=a/2+1-k-\eps$, and 
their residue cancel each others order by order 
in a way completely equivalent to what was discussed above
for $\eps\to 0$.

To conclude this section, going back to the
more convenient cutoff regularization, one may alternatively 
understand perhaps more
qualitatively the specific pole mass renormalon 
disappearance
by  examining the asymptotic behaviour of the off-shell mass
expression (\ref{GNmass}), first  expanded in powers
of $M^2/p^2$, and proceeding with the MB transform
similarly to the above described procedure. [This is then a 
completelly similar calculation than the one performed in details in 
ref.\cite{BenBraKi} for
the $O(N)$ sigma model.]
Skipping many details, we
simply indicate
sketchily that for the GN model, it would give coefficients of the 
$(M^2/p^2)^n$ terms with a form similar to Eq. (\ref{KMB}), 
but with essentially
the replacements $\mu^2 \to p^2$; $(1+s)^{-1} \to \sum_i (1+s+i)^{-1}$, $0 \le
i \le n-1$.  Those poles at $s=-1-i$, 
of residues $\sim \Gamma[t-1-i]$, produce in turn
poles at (integer) $t =1,2,..n$. Again, the integral over $t$ may be cast into
a Borel transform formally similar to (\ref{MPresidue}), 
identifying a NP part and a perturbative part. 
However, in this off-shell case, the specific poles at $t_0 =1,2,.. $ 
within perturbative terms 
of order $(M^2/p^2)^n$, are cancelled by NP term of different 
order $n+t_0$. Since individual terms are singular, one cannot
truncate the power expansion unless a definite integration contour prescription
to avoid the poles is defined\cite{BenBraKi}. In other words, the
separation between the NP and perturbative part in this 
$p^2 \neq M^2$ case is ambiguous, which means
the non Borel summability 
of the perturbative series of the general two-point function
(\ref{GNmass}) for arbitrary $p^2$, as obtained from a direct calculation,
cf. remarks in section 3 (footnote \ref{NBS}). In contrast, 
for the pole mass  
any $(M^2/p^2)^n$ terms are replaced by 1, which ``flatten" all orders of
the $M^2/p^2$ expansion, so that the
different cancellations of $t > 0$ poles now occur all at once. 
\section{Conclusion and perspectives for QCD?}  
In this paper we have exhibited in details the non-trivial
disappearance of the ``naive" perturbative IR renormalons 
of the GN mass gap at order $1/N$, 
implying the direct Borel summability of the 
``educated" perturbative expansion,
as defined by Eq.(\ref{genuine}). 
Given the detailed renormalon vanishing
mechanism, we are also confident that 
it should work similarly beyond $1/N$
order, though an explicit check has not been attempted.
Concerning strictly our $1/N$ results for the GN pole mass, 
a number of additional remarks may be useful.\\
-First, we should perhaps again emphasize that the
result Eq.(\ref{genuine}) may be not a
surprising one, as it could have been easily extracted
from previous analysis of e.g. refs.~\cite{CampRoss,GN1N}, 
if not explicitly
displayed there. As motivated in introduction, our main purpose 
here was to illustrate in a calculable model how the NP contributions 
to the pole mass organize to eliminate completely the renormalon
artifacts, even though the latter are 
somewhat artificially introduced by defining 
the naive perturbation
theory, similar to the perturbative expansion of
QCD quark pole masses.\\
-Second, the (would-be)  
perturbative ambiguities due to IR renormalons for the pole mass
here discussed are severe, being of the same (${\cal O}
(\Lam)$ order than the pole mass itself.
In contrast, there are other related quantities
of interest, like most typically the GN model 
fermion condensate, where the leading order potentially dangerous
renormalon singularities in the perturbative part 
disappear\cite{DavHam}
by explicit cancellations due to the chiral symmetry,
as reexamined briefly in the end of section 3.
There, the cancellation of renormalons is 
such that the actual perturbative ambiguities are 
of higher order than the condensate dimension, thus power suppressed
and to be considered harmless. Though the latter
results are not unrelated
with the present study, note that the IR renormalon 
cancellation mechanism as we exhibited here is of a rather different 
nature, since {\em all} (but not only the leading) IR renormalon 
singularities are annihilated each other
between the purely perturbative and NP contributions, see 
e.g. Eq. (\ref{MPresidue}).
But we stress again that these results are
very specific to the pole mass. 

It is thus tempting to speculate briefly on the
possibility of ultimate (NP) renormalon disappearance
 in QCD similarly (to some extent) to
the mechanism discussed here. QCD
in the massless quark limit also has 
a mass gap, since the approximate chiral
symmetry of the light quark sector
is dynamically broken. Now {\em if} we assume 
as usual\cite{renormalons}
that the dominant IR renormalon
contributions to the pole mass is given by
graphs of a topology analogous to the first one 
in Fig 1, but with a gluon propagator 
(wavy line) dressed now with massive, constituent quarks of
mass  $M_Q \sim \Lam_{QCD} =\bar \Lam$,  one may expect this propagator
to behave in the infrared  in a way similar to Eq.(\ref{Gq2}). (Assuming also
that such quark loops are complemented 
with appropriate QCD gauge sector contributions, so to match the correct beta
function in a gauge invariant way,
as indeed usually assumed in most QCD infrared renormalon
issues\cite{renormalons}). We see no reason why such assumption
on the NP behaviour would not be consistent with the usual perturbative
behaviour, and in particular with the standard heavy quark
pole mass renormalon picture\cite{MPren}, for which $M_Q \gg \bar \Lam$. This
is also irrespective of the fact that confinement in QCD ultimately makes the
pole quark mass relevance somewhat elusive. What can be still
theoretically relevant would be to have in this way a procedure to evaluate
the NP contributions to  the light constituent quark masses $M_Q \sim \bar
\Lam$ from  first principles (and perhaps more interestingly for the NP
order parameters related to chiral symmetry breaking).
A detailed QCD analysis is however beyond the present scope and 
left for future work.\\ 
Let us finally mention that, indeed,  
there exist more ``direct" ways of modifying the 
asymptotic properties of the perturbative
expansion of e.g. the mass gap, generically in asymptotically 
free models. It relies only
on the purely perturbative information, but is based on a modification of
the usual perturbative series. Such a method\cite{Bvarconv} appears to bypass
the explicit disappearance of renormalons here exhibited, by 
directly removing the perturbative renormalon divergences, at least
for adequate ranges of an (alternative) 
perturbative expansion parameter values compatible with
the chiral limit.
\appendix
\section{Asymptotic behaviour for $M_P\gg \Lam$}
As mentioned in section 3 the use of the mass gap relation Eq.(\ref{MPdef}),  
strictly valid only at ${\cal
O}(1/N)$ and in the chiral limit (Lagrangian mass $m\to 0$), plays in fact no 
particular role in the asymptotic
properties Eq.(\ref{genuine}), of the exact $1/N$ pole mass. Consider
Eq.(\ref{r1pole}), but now for $M_P \gg \Lam$ (which illustrates e.g.
the massive theory case with $M_P \sim m \gg \Lam$): it takes the form
\be
\ds
r_1 = \int d\phi 
[e^{-\frac{\phi}{2}} \cosh \frac{\phi}{2} 
-\frac{\phi}{2}(1+\ln(\frac{M_P}{\Lam}))^{-1}] 
(\ln (\frac{M_P^2}{\Lam^2})+\frac{\phi}{\tanh
\frac{\phi}{2}})^{-1} 
\label{r1largeM}
\ee 
to be expanded in powers of $\ln^{-1}(M_P^2/\Lam^2)$,
where we used again $\phi = \ln[(\zeta+1)/(\zeta-1)]$.
Each coefficient of such an expansion contains a logarithmic divergence, 
renormalizable by subtraction
(removing essentially the divergent $(1+\phi)/2$ piece in the bracket 
of Eq.~(\ref{r1largeM})).
After some algebra we find finally that
the renormalized series has the  
leading asymptotic behaviour
\be
r_1^{(as)}(M_P \gg \Lam) 
\sim \sum^\infty_{p=0} (-1)^{p+1} \frac{p!}
{\ln^{p+1}  (\frac{M_P^2}{\Lam^2})} 
\label{r1MPlarge}
\ee

which qualitatively 
agrees asymptotically with Eq.~(\ref{genuine}), provided 
one identifies
$M_P \sim \mu \gg \Lam$, i.e. $\ln (M^2_P/\Lam^2) \sim 1/g$. 
The sign alternation in (\ref{r1MPlarge}),
leading to Borel summability, again makes the main difference 
with the behaviour obtained starting from the 
naive perturbative analysis of section 3.
\section{IR renormalons (dis)appearance in dimensional
regularization}
In dimensional
regularization (DR) with $D=2-\eps$ (so that the GN model
is renormalizable for $Re[\eps]>0$),
Eq. (\ref{r1pole}) may be cast in the form (omitting
once more the tadpole graph to simplify):
\bea
r_1 = \frac{g N}{8\pi}  (\frac{\mu}{M})^\eps
\int \frac{d q^2}{M^2}  (\frac{q^2}{M^2})^{-\eps/2}
(1-\zeta)\; G[q^2,M^2,\eps]
\label{r1eps}
\eea
where $\zeta \equiv (1+4M^2/q^2)^{(1/2)}$,
and the DR regularized auxiliary 
field propagator    
\be
G[q^2M^2,\eps]= 
\left[1-\frac{g_0 N}{2\pi}(1-\eps) \Gamma[\eps/2] \int^1_0 dx 
(M^2+q^2 x(1-x))^{(-\eps/2)} \right]^{-1} 
+{\cal O}(1/N)\;
\label{Gq2eps}
\ee
has the $\eps \to 0$ limit
Eq. (\ref{Gq2}) (of course after coupling renormalization $g_0 = Z_g g$,
which details  are not explicitly displayed, 
absorbing the divergence in $\Gamma[\eps/2]$).
Skipping the detailed calculation and unessential
overall factors, the important point is that
the large $q^2$ behaviour of Eq. (\ref{Gq2eps}) in DR,
equivalent 
of Eq. (\ref{r1pert}), replaces integrals
over $\ln q^2$ 
with integral of the form\footnote{Note that, even if DR regularizes the
momenta integrals, we define the integral over the IR region $0 < q^2 <\mu^2$
since it is the relevant one for IR renormalons.} 

\be
\sum_{a \ge -1}\sum_l
\int_0^{\mu^2} \frac{dq^2}{M^2} (\frac{q^2}{M^2})^{a/2
-\eps/2-\eps \, l/2}
\label{basinteps}
\ee  
(where the sum over $a$ comes from the expansion of $1-\zeta$,
similarly to Eq. (\ref{zeta}), while the sum over $l$ simply
comes from the geometic expansion of Eq. (\ref{Gq2eps})).
Though integration of (\ref{basinteps}) for finite
$\eps$ does not lead to factorial divergences unlike
its $\eps\to 0$ equivalent Eq. (\ref{r1pert}), it gives poles at
$1+a/2 -\eps(1+l)/2$ to be evaluated by taking  $\eps$ complex,
and the 
$\eps\to 0$ result differs for 
$Im[\eps] >0$ or $Im[\eps] <0$. This is similar to
general results\cite{David,renormalons} on
the way in which renormalon ambiguities manifest themselves
in DR. Going now to Mellin and Borel transform as
examined for the cutoff regularization in section 4,
the DR equivalent of Eq. (\ref{MPresidue})
 takes the form~\footnote{To simplify 
without much loose of generality
we can in fact neglect for our purpose in the Mellin-Borel
transform the detailed $\eps\, l/2$
dependence of the auxiliary field propagator in Eqs.
(\ref{Gq2eps}), (\ref{basinteps}).}
\bea
\ds
r^{DR}_1 &\sim &\int_0^\infty dt \sum_{a \ge -1}
\left[\frac{\Gamma[3+a-\eps]\Gamma[t-\eps/2-1-a/2]}
{\Gamma[2+t+a/2-\eps/2]}  
\right. \nn \\ 
& & \left. +
e^{-t/g} \sum^\infty_{k=0} \frac{(-1)^k \Gamma[1+2t+2k]} {k!  
\Gamma[1+2t+k]
(1+a/2-t-k-\eps/2)}(\frac{M^2}{\mu^2})^{k-a/2-1+\eps/2}  \; \right]  
\label{MPreseps}
\eea
where the first and second term in bracket occur
as residues of poles at $s=-1 +\eps/2$
and $s=a/2-t-k$, respectively, where $s$ is the Mellin transform
variable, cf. Eq. (\ref{Mellin}). Both terms
have poles at $t=a/2+1-k-\eps/2$, and 
their residue cancel each others order by order 
in a way completely similar to Eq. (\ref{MPresidue}) for $\eps\to 0$.   

\end{document}